\documentclass[twocolumn,prd,aps,amsfonts,showpacs,preprintnumbers]{revtex4}

\newcommand{\be}{\begin{eqnarray}}
\newcommand{\ee}{\end{eqnarray}}
\newcommand{\nn}{\nonumber}

\newcommand{\Oo}{\mathtt{O}}
\newcommand{\reals}{\mathbb{R}}
\newcommand{\ints}{\mathbb{Z}}
\newcommand{\oct}{\mathbb{O}}
\newcommand{\ti}{\text{i}}

\begin{document}

\preprint{AEI-2009-062}
\preprint{ULB-TH/09-23}

\title{Supersymmetric quantum cosmological billiards}

\author{Axel Kleinschmidt}
\affiliation{Physique Th\'eorique et Math\'ematique \& International Solvay Institutes, 
Universit\'e Libre de Bruxelles, Boulevard du Triomphe, ULB-CP231, BE-1050 Bruxelles, Belgium }

\author{Michael Koehn}
\author{Hermann Nicolai}
\affiliation{Max-Planck-Institut f\"ur Gravitationsphysik, Albert-Einstein-Institut, 
Am M\"uhlenberg 1, DE-14476 Golm, Germany}

%\date{\today}

\begin{abstract}
$D=11$ Supergravity near a space-like singularity admits a cosmological 
billiard description based on the hyperbolic Kac--Moody group $E_{10}$. 
The quantization of this system via the supersymmetry constraint is shown 
to lead to  wavefunctions involving automorphic (Maass wave) forms under the 
modular group $W^+(E_{10}) \cong PSL_2(\Oo)$ with Dirichlet boundary 
conditions on the billiard domain. A general inequality for the Laplace 
eigenvalues of these automorphic forms implies that the wave function of 
the universe is generically complex and always tends to zero when approaching 
the initial singularity. We discuss possible implications of this result for the 
question of singularity resolution in quantum cosmology and
comment on the differences with other approaches. 
\end{abstract}

\pacs{04.60.-m, 04.60.Kz, 04.65.+e, 98.80.Qc}
\maketitle

%%%%%%%%%%%%%%%%%%%% Text starts here

One of the surprising results in the study of classical gravity very close 
to a generic space-like singularity due to Belinskii, Khalatnikov and 
Lifshitz (BKL)~\cite{BKL} was the realization that, under rather general 
assumptions, the system becomes ultralocal in space and can be described 
by a sequence of Kasner regimes. In the strict limit towards the singularity, 
the Kasner behavior is interspersed with hard reflections of the logarithms 
of the spatial scale factors off infinite potential walls~\cite{CWM,RS}.
This behavior has been termed `cosmological billiards'; the
geometry of the billiard table as well as the possible occurrence of 
chaotic oscillations near the singularity depend on the dimension 
and the matter content of the theory~\cite{CosmoBilliards,LivRev}. 
In particular, for $D=11$ 
supergravity it was shown by Damour and Henneaux \cite{Damour:2000hv} that 
the billiard domain is the fundamental Weyl chamber $\mathcal{C}$ of the 
hyperbolic Kac--Moody group $E_{10}$, implying chaotic behavior.

Resolving the cosmological singularity requires the transition to the 
quantum theory. In this paper we address this issue in the framework of 
cosmological billiards. More precisely, we will set up and solve the quantum 
constraints for $D=11$ supergravity for the ten spatial scale factors 
and the fermionic degrees of freedom in compliance with the supersymmetry 
constraint. The approach followed here is thus a variant of  `minisuperspace' 
quantization of gravity pioneered in \cite{Minisuperspace,Misner,Misner:1972js}
and further developed in 
\cite{Macias:1987ir,Graham:1990jd,Moniz:1996pd,D'Eath:1996at,Csordas:1995kd,Kiefer}. 
An essential new ingredient of the present work is the 
{\em arithmetic structure} provided by $E_{10}$ and its Weyl group, whose 
relevance in the context of Einstein gravity was pointed out and explored in 
\cite{Forte:2008jr}. In accordance with the M-theory proposal 
of \cite{Damour:2002cu}, where a correspondence was established at the 
classical level between a (truncated) gradient expansion of the $D=11$ 
supergravity equations of motion and an expansion in heights of roots of 
a constrained `geodesic' $E_{10}/K(E_{10})$ coset space model, 
the cosmological billiards approximation  corresponds to the restriction 
of the coset model to the Cartan subalgebra of $E_{10}$. Our results therefore represent
the first step towards the quantization of the full coset model.

Starting with the bosonic variables, the diagonal metric considered for the 
cosmological billiard for a $(d+1)$-dimensional space-time is of the 
form~\cite{dimcomment} (for $d\geq 3$)
\be
ds^2 = -N^2 dt^2 + \sum_{a=1}^{d} e^{-2\beta^a} dx_a^2\,,
\ee
leading to the kinetic term $\mathcal{L}_{\text{kin}} = \frac12 n^{-1} 
\sum_{a,b=1}^{d} \dot{\beta}^a G_{ab} \dot{\beta}^b$ in terms of the 
lapse $n=N g^{-1}$ (the spatial volume is $g^{1/2}=\exp [-\sum_a\beta^a]$) 
and the Lorentzian DeWitt metric~\cite{GK}
\be
\dot{\beta}^a G_{ab} \dot{\beta}^b \equiv 
\sum_{a=1}^{d} (\dot{\beta^a})^2 - 
\left(\sum_{a=1}^{d} \dot{\beta}^a\right)^2\,.
\ee
It will be essential that for $d=10$ this metric coincides with the restriction 
of the Cartan--Killing metric of $E_{10}$ to its Cartan subalgebra. 
The spatial ultralocality of the BKL limit reduces the 
gravitational model to a classical mechanics system of a relativistic
billiard ball described by the $\beta^a$ variables moving on straight 
null lines in the Lorentzian space with metric $G_{ab}$ until hitting 
a billiard table wall. The straight line segments are Kasner regimes. 
The singularity is at $t=+\infty$ in the `Zeno-like' time co-ordinate $t$ that is related 
to physical (proper) time $T$ by $t\sim-\log T$. There is one such system for each spatial point ${\bf x}$, 
and these systems are all decoupled.

The conjugate canonical bosonic variables of the billiard system are 
$\beta^a$ and $\pi_a = G_{ab}\dot{\beta}^b$; the Hamiltonian then is
$\mathcal{H}=\frac12\pi_a G^{ab} \pi_b$ with the inverse metric $G^{ab}$, and we have set the lapse $n=1$. 
Before quantisation, we perform the following change of variables  
by means of which the billiard motion can be projected onto the unit 
hyperboloid in $\beta$-space \cite{CosmoBilliards}
\be
\beta^a = \rho \omega^a\,,\quad \omega^a G_{ab} \omega^b =-1\,,\quad 
\rho^2=-\beta^aG_{ab}\beta^b\,,
\ee
where $\rho$ is the `radial' direction in the future light-cone and 
$\omega^a=\omega^a(z)$ are expressible as functions of $d-1$ coordinates 
$z$ on the unit hyperboloid. The limit towards the singularity is 
$\rho\to\infty$ in these variables. The Wheeler-DeWitt (WDW) operator 
on $\beta$-space takes the form~\cite{Ordering}
\be\label{WDWHam}
\mathcal{H} \equiv G^{ab}\partial_a \partial_b = -\rho^{1-d}\frac{\partial}{\partial\rho}\left(\rho^{d-1}\frac{\partial}{\partial\rho}\right) + \rho^{-2} \Delta_{\text{LB}}\,,
\ee
where $\Delta_{\text{LB}}$ is the Laplace--Beltrami operator on the 
$(d-1)$-dimensional unit hyperboloid. The WDW equation therefore 
reads $\mathcal{H} \Phi(\rho,z)=0$ 
for the wavefunction $\Phi(\rho,z)$. As usual 
(see e.g. \cite{Kiefer}) one can adopt $\rho$ as a time coordinate in the 
initially `timeless' WDW equation, with the standard 
(Klein--Gordon-like) invariant inner product 
\be\label{KGProduct}
(\Phi_1,\Phi_2) = i \int d\Sigma^a 
\Phi_1^* \stackrel{\leftrightarrow}{\partial_a} \Phi_2
\ee
where the integral is to be taken over a spacelike hypersurface inside
the forward lightcone in $\beta$-space.

In order to construct solutions we separate variables by means of the ansatz
$\Phi(\rho,z)= R(\rho)F(z)$ \cite{Misner:1972js,Graham:1990jd}. 
For any eigenfunction $F (z)$ obeying
\be\label{Laplace}
-\Delta_{\text{LB}} F (z) = E F(z)
\ee
the associated  radial equation is solved by
\be\label{R}
R_\pm(\rho) = \rho^{-\frac{d-2}{2}} 
e^{ \pm i  \sqrt{E-\left(\frac{d-2}{2}\right)^2}\log\rho}\,.
\ee
Positive frequency waves emanating from the singularity correspond to $R_-(\rho)$ and have positive inner product (\ref{KGProduct}).
To study the eigenvalues of the Laplace--Beltrami operator on the unit 
hyperboloid we use a generalized upper half plane model 
$z=(\vec{u},v)$ for the unit hyperboloid 
with co-ordinates $\vec{u}\in \reals^{d-2}$ and $v\in \reals_{>0}$ 
and the Poincar\'e metric $ds^2 = v^{-2}(dv^2 + d\vec{u}^{\, 2})$ such that
\be\label{DLB}
\Delta_{\text{LB}} = v^{d-1} \partial_v\left(v^{3-d}\partial_v\right) 
+ v^2 \partial_{\vec{u}}^2\,.
\ee
For the spectral problem we must specify boundary conditions. For the
cosmological billiard, these are provided by infinite (`sharp') potential 
walls which encapsulate the effect of spatial inhomogeneities
and matter fields near the spacelike singularity, as explained in 
\cite{CosmoBilliards,LivRev}. Following the original suggestion of 
\cite{Misner:1972js}, we are thus led to impose the vanishing of the 
wavefunction on the boundary of the fundamental domain specified by these 
walls. Accordingly, let $F(z)$ be any function on the hyperboloid 
satisfying (\ref{Laplace}) with {\em Dirichlet conditions}  at the 
boundaries of this domain~\cite{Neumann}. A direct generalization 
of the arguments on page~28 of Ref.~\cite{Iwaniec} gives
\be
-(\Delta_{\text{LB}}F , F ) 
\geq \int dv\, d^{d-2}u\, v^{3-d} (\partial_v F)^2\,
\ee
with (\ref{Laplace}) and (\ref{DLB}). Considering also
\be
(F,F) &=& \int dv\,d^{d-2}u\, v^{1-d} F^2\nn\\
& =& \frac2{d-2} \int dv\, d^{d-2}u\,v^{2-d} F \partial_v F\,,
\ee
the Cauchy--Schwarz inequality entails
\be\label{eigenvalues}
E \geq \left(\frac{d-2}{2}\right)^2\,.
\ee
From the explicit solution (\ref{R}) we thus conclude that $R_\pm(\rho)\to 0$ 
when $\rho\to\infty$, and therefore {\em the full wavefunction and all its 
$\rho$ derivatives tend to zero near the singularity}. While this conclusion
also holds with Neumann boundary conditions (for which $E\geq 0$), the above
inequality furthermore ensures that the full wavefunction is generically
complex and oscillating.

Let us now return to maximal supergravity, where the wavefunctions can
be further analyzed by exploiting the underlying symmetry encoded in the 
Weyl group $W(E_{10})$ and its arithmetic properties, and in particular 
the new links between hyperbolic Weyl groups and generalized modular groups
uncovered in \cite{Feingold:2008ih}. The Weyl reflections that the classical 
particle is subjected to when colliding with one of the walls are norm preserving,
and therefore the reflections can be projected to any hyperboloid 
of constant $\rho$, inducing a non-linear action on the co-ordinates 
$z$ (given in (\ref{wz}) below for the fundamental reflections).
For physical amplitudes to be invariant under the Weyl group, the full 
wavefunction (in $\beta$-space) must transform as follows 
\be\label{wPhi}
\Phi(\rho,z) = \pm \Phi(\rho,w_I\cdot z)
\ee
for the ten generating fundamental reflections $w_I$ of $W(E_{10})$, labeled by $I=-1,0,1,\ldots,8$. Restricting
the wavefunction to the fundamental Weyl chamber, one easily checks that
the plus sign in (\ref{wPhi}) corresponds to Neumann boundary conditions, 
and the minus sign to Dirichlet conditions (which we adopt here).
$\Phi(\rho,z)$ is thus invariant under even Weyl transformations 
$s\in W^+(E_{10})$ irrespective of the chosen boundary conditions.

Choosing coordinates as in (\ref{DLB}) the relevant variables
now live in a $9$-dimensional `octonionic upper half plane' with $z=u+\ti v$ 
where $u\equiv\vec{u}\in\oct$ is an {\em octonion}. The ten fundamental 
reflections of $W(E_{10})$ act as $(j=1,\dots,8)$
\be\label{wz}
w_{-1}(z)= \frac1{\bar{z}} \,,\,
w_0 (z) = - \theta\bar{z}\theta + \theta \,,\,
w_j (z) = -\varepsilon_j\bar{z}\varepsilon_j
\ee
where $\bar{z}:= \bar{u}-\ti v$, with $\ti u=\bar{u}\ti$ in accordance 
with Cayley--Dickson doubling \cite{Conway}. $\varepsilon_j$ and $\theta$, 
respectively, are the simple roots and the highest root of $E_8$ expressed 
as unit octonions~\cite{Conway,Feingold:2008ih}. We note that for 
$u\in\reals$ the formulas (\ref{wz}) reduce to the ones familiar from 
complex analysis, with $z\mapsto 1/\bar{z}$, $z\mapsto -\bar{z}+1$ and 
$z\mapsto -\bar{z}$, generating the group $PGL_2(\ints)$. For
{\em even} Weyl transformations, we re-obtain the standard modular 
group $PSL_2(\ints)$ generated by $S(z)=(w_{-1}w_1)(z)= -1/z$ and
$T(z)=(w_{0} w_1)(z) = z+1$. Similarly, the even Weyl group $W^+(E_{10})$ 
is isomorphic to $PSL_2(\Oo)$ where $\Oo$ are the integer octonions
(`octavians', see \cite{Conway}) and the group $PSL_2(\Oo)$ is 
{\em defined} by iterating the action of (\ref{wz}) an even number 
of times \cite{Feingold:2008ih}. Consequently, for maximal supergravity 
the bosonic wavefunctions $\Phi(\rho,z)$ 
are {\em odd Maass wave forms} for $PSL_2(\Oo)$, that is, invariant 
eigenfunctions of the Laplace--Beltrami operator transforming with a
minus sign in (\ref{wPhi}) under the extension $W(E_{10})$ of $PSL_2(\Oo)$.
Understanding these modular functions remains an outstanding mathematical 
challenge, see \cite{Eie:1992} for an introduction (and \cite{Iwaniec} 
for the $PSL_2(\ints)$ theory). For the groups $PSL_2(\ints)$ and 
$PSL_2(\ints[i])$ the (purely discrete) spectra of odd Maass wave forms 
have been investigated numerically in 
\cite{Steil,Hejhal,Bogomolny,Then2,Aurich:2004ik}. 
By modular invariance, the wavefunctions can be restricted 
to the fundamental domain  of the action of $W(E_{10})$, and conversely,
their modular property defines them on the whole hyperboloid. 
The Klein--Gordon inner product (\ref{KGProduct}) must likewise 
be restricted to the fundamental chamber 
\be
(\Phi_1, \Phi_2) = i \int_{\mathcal{F}} d {\rm vol}(z)  
\rho^{d-1} \Phi_1^* \stackrel{\leftrightarrow}{\partial_\rho} \Phi_2\,.
\ee
where $\mathcal{F}$ is the intersection of $\mathcal{C}$ with the unit 
hyperboloid. This is the canonical quantum gravity analog of the one-loop 
amplitude in string theory which is rendered finite upon `division' by the 
modular group $PSL_2(\ints)$.

We now turn to the extension of the quantum billiard analysis to maximal supergravity and 
restrict to $d=10$ henceforth \cite{Damour:2005zs,de Buyl:2005mt,Damour:2006xu}.
Classically, the gravitino $\psi_\mu$ of $D=11$ supergravity performs a 
separate fermionic billiard motion~\cite{Damour:2009zc}. This 
is most easily expressed in a supersymmetry gauge $\psi_t =\Gamma_t 
\Gamma^a\psi_a$ \cite{Damour:2005zs} and in the 
variables \cite{Damour:2009zc} (with $\Gamma_*=\Gamma^1\cdots\Gamma^{10}$)
\be\label{BilliardFermions}
\varphi^a = g^{1/4}  \Gamma_* \Gamma^a \psi^a   
\quad\text{(no sum on $a=1,\ldots,10$)}\,.
\ee
Using (\ref{BilliardFermions}) the Dirac brackets between two gravitino 
variables (see (6.3) in \cite{Damour:2006xu}) become 
$\{ \varphi^a_\alpha,\varphi^b_\beta\} = -2i G^{ab}
\delta_{\alpha\beta}$, where we have written out the $32$ real spinor 
components using the indices $\alpha, \beta$. The fermionic and bosonic 
variables are linked by the supersymmetry constraint
\be\label{susycon}
\mathcal{S}_\alpha \equiv \sum_{a,b=1}^{10} \dot{\beta}^a G_{ab} 
\varphi^b_\alpha = \sum_{a=1}^{10} \pi_a \varphi^a_\alpha =0\,.
\ee
The supersymmetry constraint implies the Hamiltonian constraint 
$\mathcal{H}=0$ by closure of the algebra
\be
\frac14\left\{ \mathcal{S}_\alpha , \mathcal{S}_\beta \right\} = \delta_{\alpha\beta} \mathcal{H}\,.
\ee
In order to quantize this system we rewrite the 320 real gravitino 
components $\varphi^a_\alpha$ in terms of 160 complex ones, and replace 
the Dirac brackets by canonical anticommutators to obtain a fermionic 
Fock space of dimension $2^{160}$ over the vacuum $|\Omega\rangle$.
For the supersymmetry constraint this amounts to the redefinition 
$\tilde{\mathcal{S}}_A = \mathcal{S}_A+ i \mathcal{S}_{A+16}$ for
$A = 1,...,16$. The quantum constraint is then solved by
\be\label{formalsol}
|\Psi\rangle = \prod_{A=1}^{16}\tilde{\mathcal{S}}^\dagger_A 
          \Big( \Phi(\rho,z) |\Omega\rangle \Big) \,,
\ee
(with $\tilde\mathcal{S}_A|\Omega\rangle =0$) 
if and only if the function $\Phi(\rho,z)$ is a solution of the 
WDW equation $\mathcal{H}\Phi=0$.
While this solution is close to the `bottom of the 
Dirac sea', there is an analogous one `close to the top' with 
$\tilde{\mathcal{S}}^\dagger_A$ replaced by $\tilde{\mathcal{S}}_A$ 
and $|\Omega\rangle$ by the completely filled state.

The cosmological billiards description is very useful but takes into account
the dependence on spatial inhomogeneities and matter degrees of freedom
only in a very rudimentary way via the infinite potential walls. It would
thus be desirable to develop an approximation scheme for the quantum state
in line with the `small tension' expansion proposed in \cite{Damour:2002cu}, 
and thereby hopefully resolve the difficulties encountered in extending 
the `dictionary' of \cite{Damour:2002cu} to higher order spatial gradients 
and heights of roots in a quantum mechanical context. In the BKL approximation,
the full wavefunction is expected to factorize as
\be\label{FullPsi}
|\Psi_{\text{full}}\rangle \sim \prod_{\bf x} |\Psi_{\bf x}\rangle\,,
\ee
near the singularity into a formal product over wavefunctions of the 
type (\ref{formalsol}), one for each spatial point (with independent 
bosonic wavefunctions $\Phi_{\bf{x}}\big(\rho({\bf{x}}),z({\bf{x}})\big)$ 
and {\em space-dependent} diagonal metric variables 
$\beta^a({\bf{x}})\equiv (\rho({\bf{x}}),z({\bf{x}}))$. The task is 
then to replace the formal expression (\ref{FullPsi}) by a wavefunction 
depending on the (infinite) tower of $E_{10}$ degrees of freedom, effectively 
implementing the de-emergence of space and time near the cosmological 
singularity, and their replacement by purely algebraic 
concepts~\cite{Damour:2005zs,Damour:2008zza}. We note that
as a consequence of the uniqueness of the standard bilinear
(Cartan--Killing) form on $E_{10}$ there is 
a {\em unique} $E_{10}$ extension of the billiard Hamiltonian 
(\ref{WDWHam}) given by
\be\label{E10Ham}
\mathcal{H} \to \mathcal{H} + 
\sum_{\alpha\in\Delta_+(E_{10})}\sum_{s=1}^{\text{mult}(\alpha)} 
e^{-2\alpha(\beta)}\Pi_{\alpha,s}^2\,.
\ee
where the first sum runs over the positive roots $\alpha$ of $E_{10}$.
This extended system requires additional constraints. A first step
in this direction was taken in \cite{Damour:2007dt} where a correspondence
was established at low $E_{10}$ levels between the classical canonical
constraints of $D=11$ supergravity (in particular, the diffeomorphism and 
Gauss constraints) on the one hand, and a set of constraints that can be 
consistently imposed on the $E_{10}/K(E_{10})$ coset space dynamics on the
other. The fact that the latter can be cast in a `Sugawara-like' form
as quadratic expressions in terms of the $E_{10}$ Noether charges 
\cite{Damour:2007dt} would make them particularly amenable for the 
implementation on a quantum 
wavefunction. In addition, one would expect that $PSL_2(\Oo)$ must be 
replaced by a much larger `modular group' whose action extends beyond 
the Cartan subalgebra degrees of freedom all the way into $E_{10}$, 
perhaps along the lines suggested in \cite{Ganor}.

As noted above, the inequality (\ref{eigenvalues}) implies that  
$\Phi(\rho,z)\to 0$ for $\rho\to\infty$, and hence the wavefunction $\Psi$ 
vanishes at the singularity, in such a way that the norm is preserved. 
Its oscillatory nature entails that it cannot be analytically extended 
beyond the singularity, a result whose implications for the question 
of singularity resolution in quantum cosmology remain to be explored. 
The mechanism usually invoked to resolve singularities in canonical approaches
to quantum geometrodynamics would be to replace the classical `trajectory' 
in the moduli space of 3-geometries (that is, WDW superspace) by a quantum 
mechanical wave functional which `smears' over the singular 3-geometries. 
By contrast, the present work suggests a very different picture, namely 
the `resolution' of the singularity via {\em the effective disappearance 
(de-emergence) of space-time} near the singularity (see also 
\cite{Damour:2008zza}). The singularity would 
thus become effectively `unreachable'. This behavior is very 
different from other possible mechanisms, such as the Hartle--Hawking no 
boundary proposal \cite{Hartle:1983ai}, or cosmic bounce scenarios of the type
considered recently in the context of minisuperspace loop quantum cosmology 
\cite{Bojowald:2001xe,Ashtekar:2007em,Bojowald:2008ec}, both of which
require continuing the cosmic wavepacket into and beyond the singularity
at $\rho=\infty$.

A key question for singularity resolution concerns the role of observables, 
and their behavior near the singularity. While no observables (in the sense 
of Dirac) are known for canonical gravity, we here only remark that
for the $E_{10}/K(E_{10})$ coset model the conserved $E_{10}$ Noether
charges do constitute an infinite set of observables, as these charges
can be shown to commute with the full $E_{10}$ Hamiltonian (\ref{E10Ham}).
The expectation values of these charges are thus the only quantities 
that remain well-defined and can be sensibly computed in the deep 
quantum regime, where the $E_{10}/K(E_{10})$ coset model is expected to
replace space-time based quantum field theory.

\vspace{0.3cm}

 \acknowledgments{We thank T. Damour, C. Hillmann, A. Rendall and H. Then 
for discussions. AK is a Research Associate of the Fonds de la 
Recherche Scientifique--FNRS, Belgium.}


\begin{thebibliography}{40}
\bibitem{dimcomment} We here keep the dimension $d+1$ of space-time 
  arbitrary but will specialize to $D=11$ supergravity below.
\end{thebibliography}
\end{document}